\documentclass[twocolumn]{aastex63}

\usepackage{amsmath,amssymb}
\usepackage{enumitem}

\newcommand{\Otwo}{O$_2$}

\newcommand{\Othree}{O$_3$}

\newcommand{\flife}{$f_\text{life}$}

\newcommand{\fOtwot}{$f_\text{\Otwo}(t)$}

\usepackage{ulem}

\shorttitle{Testing Earth-like atmospheric evolution on exo-Earths}
\shortauthors{Bixel \& Apai}

\begin{document}

\title{Testing Earth-like atmospheric evolution on exo-Earths through oxygen absorption: required sample sizes and the advantage of age-based target selection}

\author[0000-0003-2831-1890]{Alex Bixel} 
\affiliation{Department of Astronomy/Steward Observatory, The University of Arizona, 933 N. Cherry Avenue, Tucson, AZ 85721, USA}
\affiliation{Earths in Other Solar Systems Team, NASA Nexus for Exoplanet System Science}

\author[0000-0003-3714-5855]{D\'aniel Apai}
\affiliation{Department of Astronomy/Steward Observatory, The University of Arizona, 933 N. Cherry Avenue, Tucson, AZ 85721, USA}
\affiliation{Earths in Other Solar Systems Team, NASA Nexus for Exoplanet System Science}
\affiliation{Lunar and Planetary Laboratory, The University of Arizona, 1640 E. University Blvd, AZ 85721, USA}

\begin{abstract}
Life has had a dramatic impact on the composition of Earth's atmosphere over time, which suggests that statistical studies of other inhabited planets' atmospheres could reveal how they co-evolve with life. While many evolutionary pathways are possible for inhabited worlds, a possible starting hypothesis is that most of them evolve similarly to Earth, which we propose could lead to a positive ``age-oxygen correlation'' between the ages of inhabited planets and the fraction which have oxygen-rich atmospheres. We demonstrate that next-generation space observatories currently under consideration could test this hypothesis, but only if the stellar age distribution of the target sample is carefully considered. We explore three possible parameterizations of the age-oxygen correlation, finding that they yield similar results. Finally, we examine how abiotic oxygen sources could affect the results, and discuss how measuring the age-dependence of oxygen could shed light on whether it is a reliable biosignature. Future efforts can expand upon this groundwork by incorporating detailed models of the redox balance of terrestrial planets and its dependence on stellar and planetary properties.
\end{abstract}


\section{Introduction} \label{sec:introduction}

The coming decades promise exciting developments in the search for life beyond Earth, with multiple groups proposing the construction of novel space observatories which could discover and characterize several potentially Earth-like planets orbiting nearby stars \citep[e.g.,][]{HabEx2019,LUVOIR2019,Apai2019,Origins2019,Staguhn2019}. By discovering biosignature molecules in these planets' atmospheres \citep[e.g.,][]{Schwieterman2018}, such observatories would enable the first constraints on the frequency of life in the universe and comparative studies of the properties of inhabited worlds.

Molecular oxygen (\Otwo), and its photochemical byproduct ozone (\Othree), have been discussed as promising biosignatures for such missions, as \Otwo\ has a short lifetime in the Earth's atmosphere and is replenished almost entirely by photosynthetic life \citep[e.g.,][]{Owen1980,DesMarais2002}. \Otwo\ would make for an even stronger biosignature if it were found in the presence of reduced gasses (such as methane) which would quickly eliminate it in the absence of a strong oxygen source \citep[e.g.,][]{Lovelock1965,Meadows2018b}. Oxygenic photosynthesis makes use of carbon dioxide, water, and light, which have been accessible on Earth throughout its history - suggesting that many extraterrestrial ecosystems may have converged on the same mechanism \citep[e.g.,][]{Meadows2017}.

However, the presence of oxygen in Earth's atmosphere has evolved over time, with the planet having an anoxic atmosphere for approximately the first half of its history. During the Hadean and Archean eras, the abundance of \Otwo\ was no more than $10^{-6}$ times its present atmospheric level (PAL) \citep{Zahnle2006,Catling2020}. Then, during the ``Great Oxidation Event'' (hereafter GOE) circa 2.4-2.1 Gya, the concentration of \Otwo\ dramatically increased to between $10^{-4}$ to $10^{-1}$ PAL, and would later increase again (circa 600 Mya) to reach its modern abundance \citep{Lyons2014}.

The precise causes and timing of the GOE are a matter of ongoing research - for a thorough review, see \cite{Lyons2014}. While the evolutionary development of oxygenic photosynthesis was a prerequisite for the GOE to occur, the two were not necessarily coeval; in fact, the evidence suggests a delay - perhaps hundreds of Myr long - between the appearance of the first organisms to produce oxygen and its eventual accumulation in the atmosphere \citep[e.g.,][]{Brocks1999,Anbar2007,Kendall2010,Kurzweil2013,Planavsky2014}. Regardless of its causes, the GOE counts among the most dramatic changes to Earth's atmosphere in geological history, and was dependent on the existence of life. Since Earth's atmosphere was anoxic for about half of its history, some authors have considered how pre-GOE Earth analogs might appear in reflected or transmitted light, and how the presence of life on such worlds could be inferred in the absence of oxygen \citep[e.g.,][]{Pilcher2003,Domagal-Goldman2011,Seager2013,Arney2016,Arney2018,Krissansen-Totton2018}.

To date, most studies of oxygen and other potential biosignatures have focused on how life could affect individual planets. However, due to the challenging nature of characterizing terrestrial planets, future space telescopes may provide only limited information about the atmospheres and fundamental parameters (e.g. bulk composition) of many individual planets. Even still, important information will be enclosed in the overall population of planets studied, and trends between their properties can be tested against the predictions of models for terrestrial planet evolution \citep[e.g.,][]{Bean2017,Checlair2019}. In such cases, sample sizes will be a limiting factor on the complexity of models which can be tested.

Since Earth's biosphere and atmosphere have co-evolved over time, the exciting possibility exists that by studying several inhabited planets spanning a range of ages, we could test for shared trends in the co-evolution of their atmospheres and biospheres, uncovering common patterns which govern the evolution of life in the universe. For example, if future space missions are able to detect the presence of \Otwo\ or \Othree\ in the atmospheres of several potentially habitable worlds, this would allow them to constrain the frequency of oxygen-bearing planets as a function of age. We propose that this measurement could be used to test whether the atmospheric evolution of other inhabited worlds resembles that of Earth (i.e. the null hypothesis). By this, we mean that they start with anoxic atmospheres which eventually become oxygen-rich - although the time required for oxygenation will likely vary between planets \citep{Catling2005}. If they evolve like Earth, then the fraction of inhabited planets with oxygenated atmospheres should increase with age, with older planets being more likely to have undergone a GOE-like event. If such a trend were discovered, it would strongly suggest that Earth-like atmospheric evolution is typical for inhabited planets, and would by extension strengthen the case for \Otwo\ as a biosignature, as we can think of no plausible abiotic explanation for this trend.

In this paper, we estimate how many potentially habitable planets a future telescopic biosignature survey must characterize to detect a positive ``age-oxygen correlation'' -- and thus test whether Earth's atmospheric evolution is typical. We present our results as a function of the actual occurrence rate of life on potentially habitable worlds, and investigate the optimal target stellar age distribution for testing our hypothesis. Finally, we estimate the noise which would be introduced by strong abiotic sources for \Otwo, and discuss how a statistical sample of planets with oxygenated atmospheres could be used to verify \Otwo\ as a biosignature.

This study does not attempt to model in detail the many possible factors affecting the rate at which inhabited planets acquire oxygen, which we discuss qualitatively in Section \ref{sec:planet_properties}. Rather, we use Earth's evolutionary history as the practical template for an initial estimate of the sample size required to begin studying their oxygen evolution. Our results should be interpreted with this caveat in mind, and we encourage future studies to build off our approach by incorporating the effects of diverse planetary parameters on the redox balance of other habitable worlds.

\section{Methods}
Here, we consider only potentially habitable planets or ``exo-Earth candidates'' (EECs). We use these terms interchangeably to refer to planets which are comparable in size to Earth and have orbits within the liquid water habitable zone, and which therefore could sustain habitable surface conditions \citep[e.g.,][]{Kasting1993,Kopparapu2013,Kopparapu2014}. Note, however, that our results are agnostic to the exact range of sizes and orbits considered to be potentially habitable, except when we compare them to the predicted discovery yields for future space-based biosignature surveys. For the reasons argued above, we also assume that inhabited planets demonstrate a positive correlation between their ages and the fraction which have oxygen in their atmospheres, which we hereafter refer to as the ``age-oxygen correlation''.

Our basic methodology is as follows: we generate a number $N$ EECs with randomly assigned ages $t$, a fraction \flife\ of which are assumed inhabited. We then assume that some age-dependent fraction \fOtwot\ of inhabited planets have undergone a GOE-like event and therefore have detectable amounts of \Otwo\ \emph{or} \Othree\ in their atmospheres. Assuming that a future space-based survey has discovered and spectroscopically characterized the entire sample, we apply a statistical test to calculate the confidence with which the age-oxygen correlation could be discovered as a function of the number of planets observed. Finally, we average the results of this test across $10^4$ random samples\footnote{In some samples none of the planets have \Otwo, so the $p$-value is undefined and we discard it. This generally only occurs for low values of $N$ and \flife\ and does not significantly impact our results.} for each cell in a two-dimensional grid of values for $N$ and \flife.{\footnote{The code used to generate the grid of p-values and Figures \ref{fig:models}-\ref{fig:alt_cases} can be found  \href{https://www.github.com/abixel/age-O2-correlation/}{here}.}

\subsection{Fraction of inhabited planets with \Otwo} \label{sec:models}
We consider three functions to describe the fraction \fOtwot\ of inhabited planets which have detectable \Otwo\ or \Othree\ in their atmospheres as a function of their age $t$ (or, approximately, the age of their host star). We consider the stellar ages to be determined precisely, and discuss the feasibility of this assumption in Section \ref{sec:determining_ages}.

The functions, plotted in Figure \ref{fig:models}, are as follows:  in the first case, \fOtwot\ increases exponentially over an e-folding timescale of 3.2 Gyr, so that the typical planet reaches its GOE at the same epoch as Earth's ($t \sim 2.2$ Gyr). In the second case, all planets evolve identically, encountering their GOEs at the same point in time as Earth did. The third case assumes the same functional form as the first, but with a longer timescale of 10 Gyr - in this case, most planets encounter their GOEs much later than Earth did. We discuss the motivation behind these functional forms in Section \ref{sec:models_discussion}.

Finally, in each case we only distinguish between oxygenated and anoxic atmospheres; the mixing ratio of \Otwo\ is not modelled. Generally speaking, the presence or absence of \Otwo\ will be much easier to determine than its precise mixing ratio. Furthermore, while \Otwo\ may not be directly detectable for planets with only a small biogenic abundance -- analogous to Earth during the Proterozoic era --  its presence might still be inferred through that of its strongly absorbing byproduct \Othree\ \citep[e.g.,][]{Angel1986,DesMarais2002,Segura2003,Reinhard2017}.

\subsection{Abiotic \Otwo} \label{sec:abiotic_O2_sources}
Several authors have suggested scenarios in which a rocky planet in or near the habitable zone could attain a detectable amount of \Otwo\ through abiotic sources, generating a ``false positive'' biosignature. A review of several of these scenarios and the means by which they can be distinguished from biological sources can be found in \cite{Meadows2018b}.

Generally speaking, abiotic oxygen is produced by the splitting of H$_2$O or CO$_2$ and the subsequent escape of hydrogen to space. Water vapor can be split either in the upper atmosphere due to UV-driven photolysis \citep[e.g.,][]{Wordsworth2014,Luger2015,Meadows2018a,Wordsworth2018} or on the surface through a photocatalytic reaction involving titanium dioxide \citep{Narita2015}. The photolysis of carbon dioxide can lead to a buildup of abiotic \Otwo, which is exacerbated in the radiative environment of low-mass stars, or when the outgassing flux of reducing species (namely H$_2$ and CH$_4$) is much lower than on Earth \citep{Hu2012,Domagal-Goldman2014,Tian2014,Gao2015,Harman2015,Hu2020}.

More optimistically, other models show that the lightning-driven recombination of CO and O \citep{Harman2018} as well as volcanic outgassing at rates comparable to Earth's \citep[e.g.,][]{Hu2012,James2018} could each counter \Otwo\ buildup from abiotic sources. Further research into such preventative factors may alleviate concerns about the reliability of this biosignature.

Some of these proposed abiotic sources of \Otwo\ could, in principle, be present on both inhabited and non-inhabited planets, and would mask the age-oxygen correlation by imbuing planets with \Otwo\ from a young age. To investigate this, we allow for some age-independent fraction of EECs to have abiotic oxygen sources, regardless of whether or not they have life. For most of the results presented below we set this parameter to zero so that there are no planets with abiotic \Otwo, but we investigate their impact in Section \ref{sec:abiotic_O2}.

\subsection{Stellar age distribution} \label{sec:ages}

Recent studies have found that while the star formation history of the Milky Way disk has varied measurably, it is to first order uniform in age, with the oldest disk stars forming $\sim 10$ Gya \citep[e.g.,][]{Snaith2015,Fantin2019,Mor2019}. Assuming homogeneous star formation throughout the disk, the age distribution of nearby stars should resemble this history, excepting a small deficit of old, massive stars with main-sequence lifetimes shorter than 10 Gyr. We therefore assume a uniform stellar and planet age distribution from 0 -- 10 Gyr for most of our results.

However, it may be more efficient to prioritize observing only young and old stars, so as to maximize the difference in \fOtwot\ for the models in Figure \ref{fig:models}. This is especially important for surveys limited by available telescope time, where the target list must necessarily be pruned. Likewise, excluding young or old stars from any survey could inhibit the detectability of the age-oxygen correlation. To examine these considerations, we also simulate samples consisting of different combinations of young (0--2 Gyr), intermediate (2--7 Gyr) and old (7--10 Gyr) age ranges. We discuss the impact of the age distribution in Section \ref{sec:age_results}.

\subsection{Correlation test} \label{sec:tests}
To determine whether the age-oxygen correlation can be confidently identified in our simulated samples, we apply two statistical tests, each implemented using SciPy \citep{Jones01}. The first is the Mann-Whitney U test \citep{Mann1947}, which is applied to measurements of a variable (i.e., age) from two independent populations (i.e., oxygenated and anoxic planets), to determine whether the age distribution of one population is stochastically greater than the other. Unlike the similar and more widely used Student's t-test, the Mann-Whitney test does not assume the two age distributions to be of normal shape and equal variance, and compares the samples through their mean ranks rather than through their sample means. We also apply Spearman's rank correlation test \citep[e.g.,][]{Wall2003}, which is usually applied to detect non-linear, monotonic correlations between two variables. The two variables can be continuous (i.e., age) or discrete (i.e., whether the planet has \Otwo).

We use the Mann-Whitney test to calculate most of our results, as we believe its underlying assumptions most accurately match our data, but we also compare its efficiency to that of Spearman's test in Section \ref{sec:test_comparison}. Each of the tests reports a $p$-value representing the probability that age and oxygen are \emph{not} apparently correlated, with values $p < 0.05$ corresponding to a significant (95\% likely) correlation. Finally, since we are testing a directional hypothesis (that age and oxygen are \emph{positively} correlated), we calculate one-tailed $p$-values.

\begin{figure*}[h]
\centering
\includegraphics[width=0.8\textwidth]{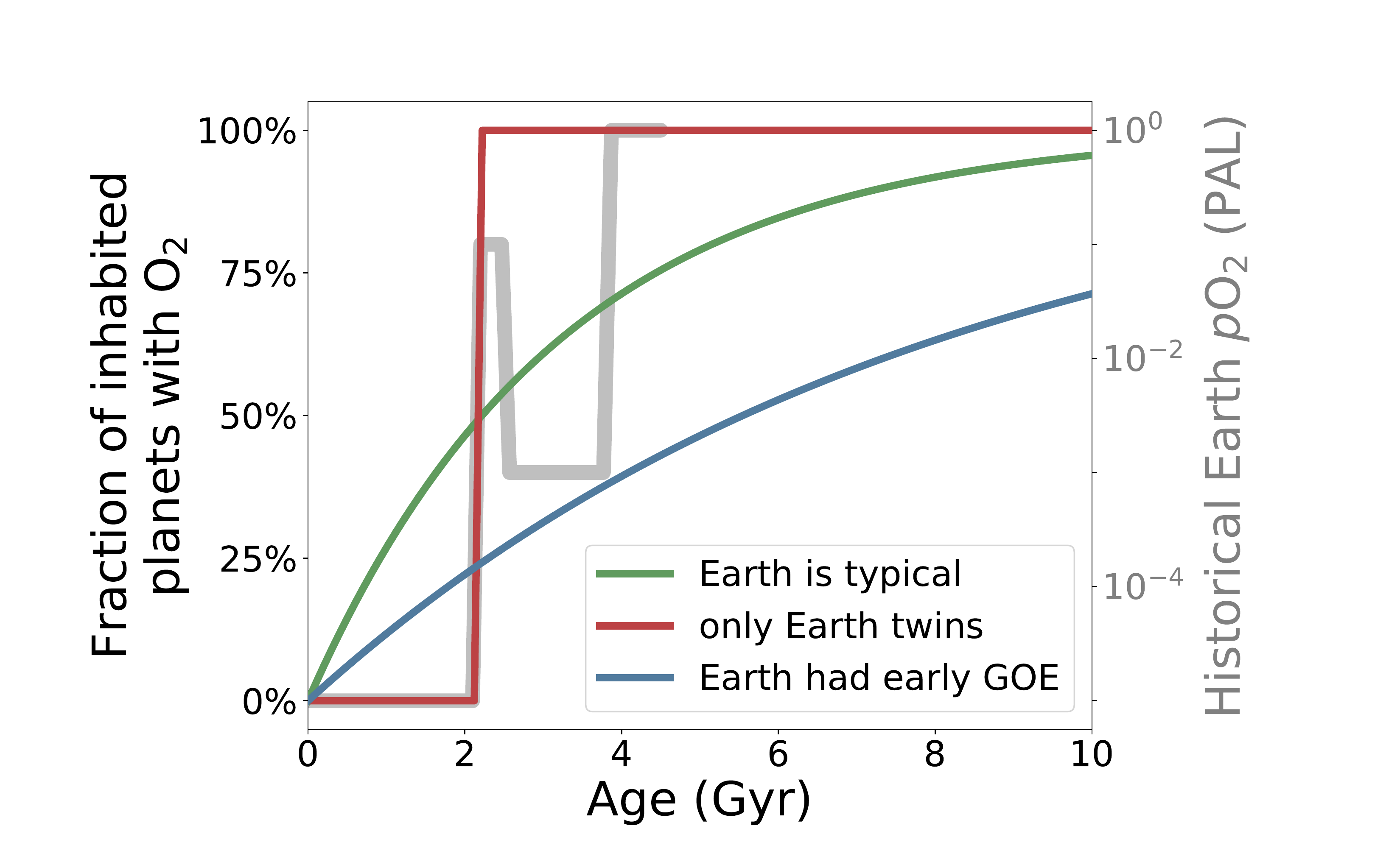}
\caption{(In color, left axis) Three functions which we assume to describe \fOtwot, the fraction of inhabited planets which have a detectable amount of oxygen as a function of age $t$. In the first two cases (green and red), the typical inhabited planet undergoes a GOE at the same epoch as Earth's ($t_H \sim 2.2$ Gyr). (In gray, right axis) We plot an estimate of Earth's historical \Otwo\ abundance versus the present atmospheric level (PAL), adapted from \cite{Reinhard2017}. Note that estimates for $p$\Otwo\ when Earth was younger than $\sim 4$ Gyr range by up to two orders of magnitude.}
\label{fig:models}
\end{figure*}

\begin{figure*}
\centering
\includegraphics[width=0.8\textwidth]{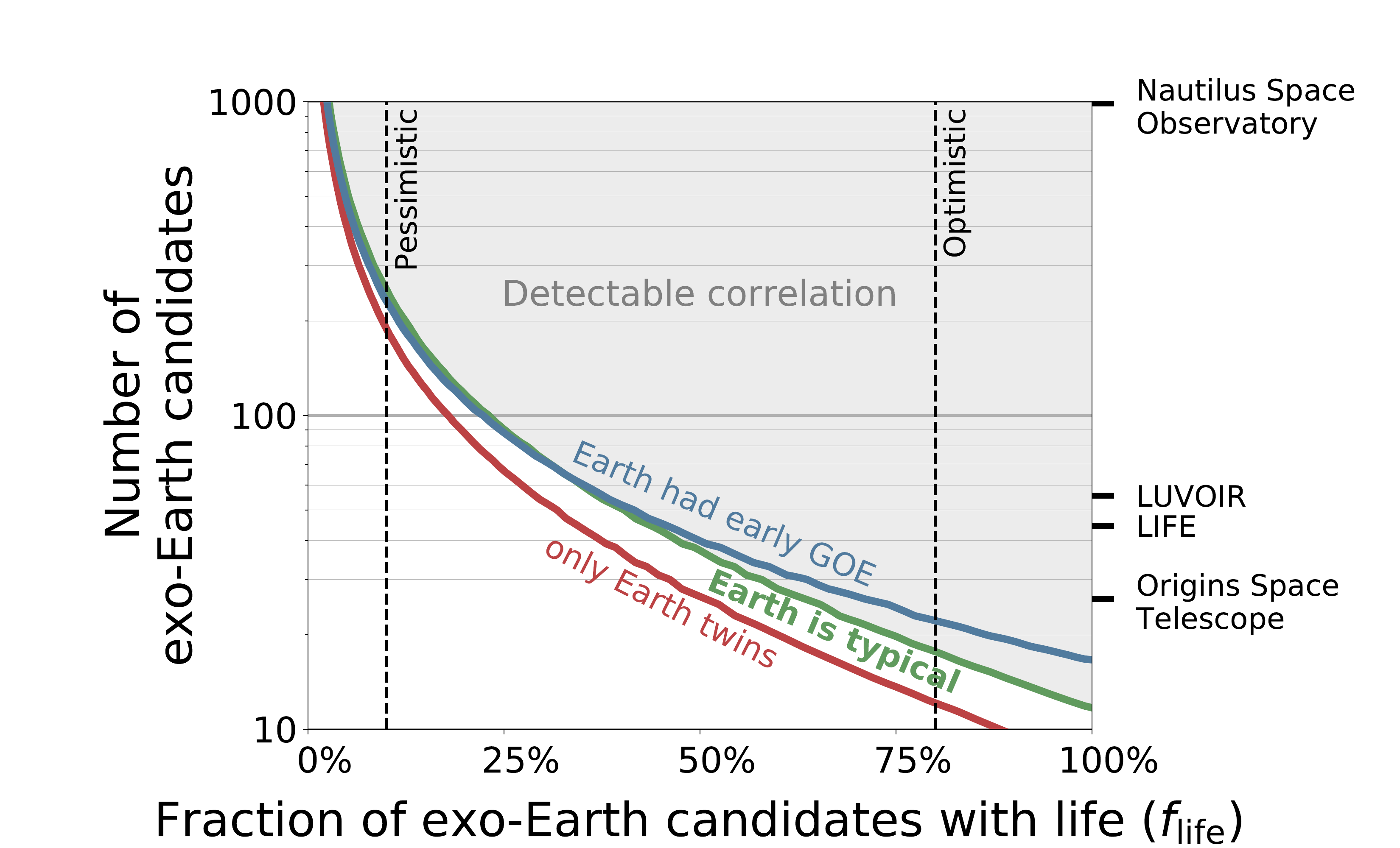}
\caption{The number of EECs as a function of \flife\ which must be characterized to confidently ($p = 0.05$) detect a correlation between their ages and the fraction with \Otwo\ or \Othree\ in their atmospheres. Results are plotted for each of the cases detailed in Figure \ref{fig:models} and a target sample with a uniform distribution of ages between 0 -- 10 Gyr. The dashed lines indicate ``pessimistic'' and ``optimistic'' cases for the frequency of inhabited worlds among EECs. On the right, we include estimates for the EEC detection yield of a few possible future observatories reviewed in Section \ref{sec:observatories}. Note, however, that these estimates were calculated using different methods, and the occurrence rates used to determine them may have been overestimated \citep[][]{Pascucci2019}.}
\label{fig:results}
\end{figure*}

\begin{figure*}
\centering
\includegraphics[width=\textwidth]{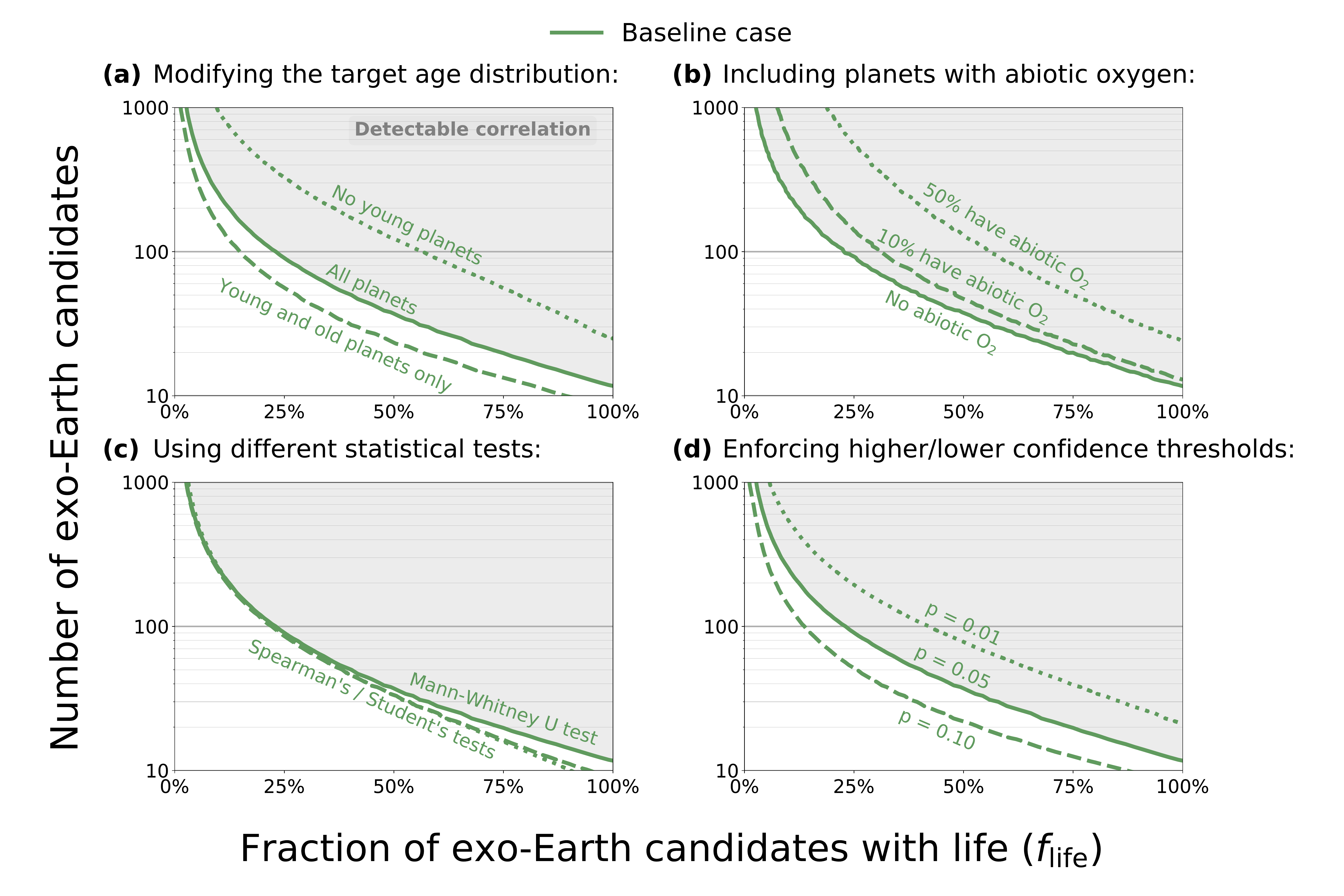}
\caption{We re-calculate the number of EECs which must be characterized to detect the age-oxygen correlation in the ``Earth is typical'' case, but under different conditions than assumed in Figure \ref{fig:results}. For comparison, we re-plot the baseline result in Figure \ref{fig:results} as a solid line in each panel. (a) We modify the age distribution of the target sample by including only young (0--2 Gyr) and old (7--10 Gyr) planets (dashed) or by excluding young planets (dotted). (b) We assume that 10\% (dashed) or 50\% (dotted) of EECs have some abiotically-produced \Otwo. This fraction includes inhabited and non-inhabited planets alike. (c) We re-calculate the results using Spearman's rank correlation test (dashed) and Student's t-test (dotted) to determine the detection significance. (d) We plot contours for a less ($p = 0.1$; dashed) or more ($p = 0.01$; dotted) confident detection of the age-oxygen correlation.}
\label{fig:alt_cases}
\end{figure*}

\section{Results}

\subsection{How many planets must be observed to detect a correlation?}
The contour plots in Figure \ref{fig:results} delineate the number of planets - as a function of \flife\ - which must be characterized to detect the age-oxygen correlation with high confidence ($p < 0.05$). The shaded area marks the region of the parameter space in which the proposed age-oxygen correlation could be identified. Each coloured contour corresponds to one of the cases in Figure \ref{fig:models} -- except where otherwise stated, we consider ``Earth is typical'' as our baseline case.

For the baseline case, we find that a sample of $\sim 20$ EECs would be sufficient to detect the age-oxygen correlation if life is present on $80\%$ of such worlds. In a more pessimistic case where life exists on only $10\%$ of these planets, then $\sim 300$ characterizations are required. If life is present on fewer than $2\%$ of EECs, then the sample size required exceeds one thousand.

\subsection{What is the optimal age distribution of target stars for this experiment?} \label{sec:age_results}

To maximize the science yield of missions capable of detecting biosignatures, it may be prudent to prioritize their targets based on age. In Figure \ref{fig:alt_cases}(a) we investigate the impact of different age-based target selection strategies on a survey's ability to test for an age-oxygen correlation.

We find that by selecting targets with ages just between 0--2 and 7--10 Gyr, the total number of planets required to detect the age-oxygen correlation drops by about 35\%. Alternatively, when a survey excludes young systems, the number of planets required increases by a factor of $\sim 3$. These results suggest that, when possible, surveys of habitable exoplanets should not prioritize intermediate and older age planets over younger targets. In fact, the youngest planets are the most important targets for testing our hypothesis, and should be prioritized in surveys which are limited by observing time rather than fundamental instrument constraints.

\subsection{What is the impact of abiotic sources for \Otwo?} \label{sec:abiotic_O2}

For most of our results, we assume that all oxygen is produced by life, but in Figure \ref{fig:alt_cases}(b) we investigate the amount of noise introduced if some age-independent fraction of EECs have abiotic \Otwo\ sources. We find the impact on the required sample size to be modest to considerable. If abiotic \Otwo\ exists on about 10\% of EECs, then the required sample size must increase by 25-100\% depending on \flife. In the pessimistic case that half of EECs have abiotic \Otwo\ and life is rare, the required sample size increases by an order of magnitude.

\subsection{What is the most efficient test for detecting the age-oxygen correlation?} \label{sec:test_comparison}
In Figure \ref{fig:alt_cases}(c) we compare results using each of the statistical tests reviewed in Section \ref{sec:tests}, as well as the more commonly-used Student's t-test (which is not strictly applicable to our case as it assumes normal age distributions). Generally speaking, the three tests yield similar results. While Student's and Spearman's tests are slightly more sensitive to the correlation for high values of \flife, the assumptions of the Mann-Whitney test most accurately match our data set, which consists of two non-normal age distributions (for oxygenated and anoxic atmospheres) with unequal variances.

While we have chosen $p = 0.05$ (i.e., a 5\% probability that age and oxygen are uncorrelated) as our threshold for a confident detection of a positive correlation, a lower or higher confidence level detection may be achieved through characterizing fewer or more EECs. In Figure \ref{fig:alt_cases}(d) we calculate the number of targets required to detect the age-oxygen correlation with low confidence ($p < 0.1$) or very high confidence ($p < 0.01$). These contours demonstrate that while a confident detection may be out of reach for a given sample size if life is rare, preliminary evidence can still be acquired to motivate a more in-depth survey.

\section{Discussion}

\subsection{Future observatories could test the proposed age-oxygen correlation} \label{sec:observatories}
Current observatories lack the capability to detect oxygen or ozone absorption in the atmospheres of terrestrial exoplanets. However, multiple ambitious space mission concepts which have been proposed in the literature could perform this characterization for statistically meaningful numbers of planets. Here, we identify several such concepts and compare their likely sample sizes (where available) to the requirements we predict in Figure \ref{fig:results}.

Two observatories would use coronagraphic instruments to directly image potential exo-Earths orbiting nearby FGK stars. The Large UV/Optical/IR Surveyor \citep[LUVOIR,][]{LUVOIR2019} would feature an 8- to 15-meter diameter segmented primary mirror, while the Habitable Exoplanet Observatory \citep[HabEx,][]{HabEx2019} would make use of a 4-meter monolithic mirror. HabEx would also launch with a starshade, which would maneuever separately from the telescope to occult the targeted host star and enable deep spectroscopic characterization of its planets. With their broad UV-to-NIR wavelength coverage, both telescopes could probe for ozone absorption at far-UV wavelengths as well as \Otwo\ in the visible spectrum. Employing the yield optimization methodology first developed by \cite{Stark2014}, both concept studies have reported estimates for the number of EECs which they could detect, namely $54^{+61}_{-34}$ for LUVOIR and $8^{+9}_{-5}$ for HabEx \citep[see also][]{Stark2015,Stark2016,Stark2016b,Stark2019,Kopparapu2018}. As such, LUVOIR would be able to test for the age-oxygen correlation given \flife\ $> 50\%$. HabEx would likely lack the sample size required to detect it on its own, but could contribute substantially towards building up a sufficiently large sample of characterized EECs if complemented by ground- or space-based efforts targeting nearby M dwarfs.

Three mission concepts would use transit and/or phase curve spectroscopy to characterize potential exo-Earths. The \emph{Nautilus} Space Observatory \citep{Apai2019} would consist of thirty-five unit space telescopes, each with an $\sim$8.5-meter diameter ultralight, diffractive-refractive lens as the primary light-collecting element. Through visible-to-NIR transit spectroscopy, the array could be used to search for oxygen or ozone in the atmospheres of up to one thousand EECs. With such a sample, \emph{Nautilus} could test the age-oxygen correlation even in the pessimistic case where only 10\% of EECs are inhabited. Two other telescopes would enable mid-infrared transit and phase curve spectroscopy of planets orbiting mid-to-late M dwarfs, including the Origins Space Telescope \citep[$\sim26$ characterized EECs,][]{Origins2019} and the Mid-Infrared Exoplanet Climate Explorer \citep{Staguhn2019}, and would infer the presence of oxygen through \Othree\ absorption between 9-10 \micron. In the optimistic case, a sample of $20-30$ planets could achieve the requirements outlined in Figure \ref{fig:results}, or it could be combined with the yield of missions targeting FGK stars.

Space-based infrared interferometry offers another avenue towards directly imaging exo-Earths through their thermal emission. The Large Interferometer For Exoplanets (LIFE) project\footnote{https://www.life-space-mission.com/} aims to coherently combine light from four $\sim 2.8$-meter mirrors in order to detect and characterize nearby exo-Earths in the 5--25 \micron\ wavelength range. LIFE would characterize the atmospheres of $\sim 45$ EECs orbiting nearby GKM stars, and with such a sample could detect the age-oxygen correlation for \flife\ $> 25\%$ \citep{Kammerer2018,Quanz2018}.

A note of caution: the yield estimates above have been adapted directly from literature references, and may not be directly comparable as they were calculated using different techniques. All of them rely on estimates of $\eta_\oplus$ (the number of potentially habitable planets per star) extrapolated from \emph{Kepler} data, but recent work by \cite{Pascucci2019} suggests that such estimates are exaggerated by a factor of 4--8$\times$, as the \emph{Kepler} radius distribution of short-period planets is heavily impacted by atmospheric loss. Nevertheless, the numbers cited suggest that testing for the age-oxygen correlation may be an achievable science goal for some of these missions.

In summary, we conclude that future space missions which are being designed to detect oxygen in the atmospheres of individual planets could use the same capabilities to test the null hypothesis that the Earth's atmospheric evolution is typical for an inhabited world. To achieve this science goal, the stellar age distribution of the target sample should be carefully considered: when it is necessary to prioritize targets, surveys are encouraged to favor young stars to maximize their sensitivity to changes in atmospheric oxygen content during the first few billion years.

\subsection{Impact of planet and stellar properties} \label{sec:planet_properties}
In our analysis we assume that the amount of time required for a planet to reach a GOE-like transition is random, but typically between $1-10$ Gyr. In reality, this timescale will be dependent on each planet's properties. In order for a GOE-like event to occur, a planet's biological oxygen source must grow large enough and/or its oxygen sinks must become minimal enough that the former overwhelms the latter. Here we address ways in which these sources and sinks might vary across different types of habitable worlds.

\cite{Catling2005} coin the term ``oxygenation time'' to refer to the time required for a planet to acquire enough atmospheric \Otwo\ to support complex life, and they argue that a planet's size, composition, and the presence or absence of continents could affect this timescale in diverse ways. For example, larger planets, or planets with more reducing initial compositions, will have a greater inventory of reducing matter to exhaust. The interior heat flux could be higher on planets larger than Earth, or for tidally-heated planets on close-in orbits around low-mass stars \citep{Driscoll2015}, which would affect outgassing rates. Continents play a role in both removing oxygen from the atmosphere (through outgassing and rock weathering) and replenishing it (through the burial of organic matter), but on a planet without continents these processes would be diminished \citep{Lunine2013}.

The escape of hydrogen from Earth's atmosphere into space acts to oxidize the planet, an effect which may have triggered the GOE since hydrogen escape from early Earth's atmosphere was likely much faster than at present \citep[][]{Catling2001,Claire2006,Zahnle2013,Zahnle2019}. On Earth, the hydrogen escape rate is limited by its diffusion rate into the exosphere \citep{Hunten1976}. The diffusion length decreases with surface gravity, which ranges from $\sim$ 0.3 -- 2.5 $g$ on rocky exoplanets \citep[e.g.,][]{Neil2020}, so diffusion-limited escape should be more efficient on larger planets - but at the same time the maximum escape rate of hydrogen from the exosphere will be throttled due to increased gravitational potential \citep{Catling2005}. In total, we might expect a non-monotonic relationship between hydrogen escape and planet size which could accelerate or inhibit the oxygenation of the atmospheres of other inhabited worlds.

Since the advent of oxygenic photosynthesis is presumed a prerequisite for global oxygenation, the rate at which evolutionary changes occur in the biosphere could also limit the oxygenation timescale. Several authors have investigated factors affecting the pace of biological evolution on planets hosted by low-mass stars, with some suggesting that prebiotic chemistry could be inhibited by a deficit of ultraviolet radiation \citep{Buccino2007,Ranjan2017,Rimmer2018}, and others proposing that complex life could take longer to subsequently evolve \citep{Haqq-Misra2018,Haqq-Misra2019}. Biogenic oxygen levels may be substantially lower around low-mass stars, as they emit less of the visible-wavelength radiation which drives oxygenic photosynthesis on Earth \citep[e.g.,][]{Kiang2007,Gale2016,Lehmer2018,Mullan2018,Ritchie2018,Lingam2018,Lingam2019}. The net effect of these factors could be to substantially delay or entirely prevent the oxygenation of the atmosphere for habitable planets orbiting low-mass stars.

A planet's size, bulk composition, and its host star's spectral type can often be constrained through observation, so it is conceivable that one could control for their effects - but this would necessitate a larger sample. It may be possible to detect continents by measuring a planet's photometric variability through extensive direct imaging observations \citep[e.g.][]{Ford2001,Cowan2009,Cowan2018,Farr2018,Lustig-Yaeger2018,Fan2019,Aizawa2020}, but likely only for a limited number of optimal targets. 

In total, we expect that the diversity of planet compositions and environments will result in a corresponding diversity of oxygenation timescales, but overall the age-oxygen correlation should remain as long as some planets were oxygenated within their first several Gyr. Our analysis accommodates this diversity by simulating planets with such timescales ranging from $\sim$ 1 -- 10 Gyr, but if the typical timescale is in fact very short ($< 1$ Gyr) or long ($>10$ Gyr), then a larger sample size will likely be required to detect the correlation. Further detailed theoretical treatments of the variation of oxygen sources and sinks across a realistic range of planetary properties will be valuable for evaluating the assumptions made in Figure \ref{fig:models}.

\subsection{Verifying \Otwo\ as a potential biosignature}

We propose that a potential future discovery of a positive age-oxygen correlation would serve as additional evidence for life on oxygen-bearing planets, even if concerns about false positives cannot be ruled out through contextual evidence for individual planets in the sample \citep[e.g.,][]{Meadows2018b}. While multiple mechanisms have been proposed for the abiotic generation of \Otwo, none so far have been shown to produce a positive correlation of \Otwo\ content with age. On the other hand, such a correlation does reflect the history of biogenic oxygen in Earth's atmosphere, so if discovered it would suggest a similar (i.e. biological) history for other worlds. Nevertheless, like all proposed biosignatures, the discovery of an age-oxygen correlation would need to be rigorously scrutinized to ensure that no plausible abiotic evolutionary scenarios could produce it.

On the other hand, if enough planets with oxygenated atmospheres are detected that a positive age-oxygen correlation can be \emph{ruled out} over Gyr timescales, this would imply that most planets which already have oxygen acquired it before $\sim$ 1 Gyr. An explanation would be required for why Earth took substantially longer to become oxygenated than most oxygen-rich planets. Such an explanation may be found through modeling the effects of planet and stellar properties on redox balance as discussed in Section \ref{sec:planet_properties}, or perhaps through an argument about anthropic bias (i.e., if complex and intelligent life is more likely to evolve on planets with late GOEs, then our planet is more likely to have had a late GOE).

\subsection{Complicating factors for detecting \Otwo\ or \Othree }

We assume that \Otwo\ or at least \Othree\ will be detectable for every post-GOE planet, but complicating factors could make this assumption optimistic. Clouds and hazes can mask absorption by low-altitude gasses, an issue which is expected to affect both direct imaging and transit observations of rocky planets \citep[e.g.,][]{Arney2017,Rugheimer2018,Wang2018,Kawashima2019,Lustig-Yaeger2019}. For transit spectroscopy, it could prove difficult to disentangle the spectral features of the atmosphere from those of the stellar photosphere \citep{Apai2018}. This may inhibit the detection of biosignatures on planets around M dwarfs \citep{Rackham2018,Zhang2018,IyerLine2019}, but would be a less prominent issue for FGK stars \citep{Rackham2019a}. 
Detailed spectral modeling of the photosphere may also help to resolve this degeneracy \citep{Pinhas2018,Rackham2019b,Wakeford2019,IyerLine2019}.

We again emphasize that even in cases where \Otwo\ is difficult to observe (e.g., due to clouds), strong \Othree\ absorption may still be visible. Generally, while these complicating factors may have the effect of increasing the sample size required to detect the age-oxygen correlation, they in principle should not inhibit it.


\subsection{Assumed correlations} \label{sec:models_discussion}
In our analysis we consider three functions to describe the fraction of inhabited planets with oxygenated atmospheres. The first and third functions have the following exponential form:
\[
f_\text{\Otwo}(t) = 1-\exp(-t/\tau)
\]
This is appropriate if we assume that every inhabited planet has an equally small probability to undergo a GOE during each consecutive interval of time during its history. If Earth is a typical example, then this probability is $\sim 3\%$ per 100 Myr. This case seems appropriate if the timing of the GOE is set by changes in the biosphere, as several random and independent evolutionary steps must occur before oxygenic photosynthesis can become a dominant form of metabolism.
 
In the second case we assume a step function:
\begin{equation}
  f_\text{\Otwo}(t) =
    \begin{cases}
      1 & \text{$t\geq 2.2$ Gyr}\\
      0 & \text{$t<2.2$ Gyr}\\
    \end{cases}       
\end{equation}
Under this case, every planet is an exact Earth analog and undergoes a GOE at the same age as did Earth. While this is unlikely to actually be correct, a step function would be the most easily detectable correlation, so our results for this case reflect a lower limit on the required sample sizes.

Despite the difference between the functional forms of the first/third and second cases, the sample sizes which must be observed to test them typically agree to within a factor of two, suggesting that our results are relatively consistent across these cases.

\subsection{Prospects for determining stellar ages} \label{sec:determining_ages}
For simplicity, we assume that each planet's age (which is approximately the age of its host star) is known with high precision, but this is not a true assumption for most known exoplanets today.

It is plausible that precise ($<1$ Gyr) age constraints could be achieved through asteroseismology, as demonstrated for several \emph{Kepler}/K2 planet hosts \citep[e.g.,][]{Mathur2012,Chaplin2014,Silva2015,Creevey2017,Kayhan2019,Lund2019}. With the same method, the PLATO mission could allow for 10\% precision age measurements of hundreds of bright solar-type stars \citep{Rauer2014}.

However, asteroseismic pulsations have so far proved difficult to detect in low-mass stars despite extensive efforts \citep{Baran2011a,Baran2011b,Krzesinski2012,Baran2013,RodriguezLopez2015,Rodriguez2016,Berdinas2017}. For now, age estimates for low-mass stars rely on various spectroscopic and photometric relations. For example, \cite{Burgasser2017} combine several diagnostics to determine the age of TRAPPIST-1, a well-studied ultra-cool dwarf known to host multiple potentially habitable planets \citep{Gillon2017}. Despite a thorough analysis, the authors are only able to constrain the age of the system with a 1$\sigma$ precision of $\pm\, 2.2$~Gyr, demonstrating that securing sub-Gyr age constraints for low-mass stars is not feasible using existing techniques.

Provided systematic errors are minimal, age uncertainties could be factored into a statistical correlation test. Doing so, however, would likely increase the required sample size if the uncertainty is much larger than the age range over which the correlation is expected ($\gtrsim 1$ Gyr). It is therefore important that advances be made over the next two decades in the measurement of stellar ages, particularly for low-mass stars.

\subsection{Luminosity evolution for low-mass stars} \label{sec:luminosity_evolution}
A potential source of uncertainty comes from the luminosity evolution of the host star, and therefore the evolution of its habitable zone. Habitable zone planets around low-mass stars are the best targets for characterization through transit spectroscopy because they produce larger relative transit depths, are more likely to transit, and transit more often than Earth twins around Sun-like stars. However, their host stars' habitable zones contract significantly during the pre-main sequence phase, which can last for hundreds of Myr. Planets towards the inner edge may have regained their habitability only after the star reached the main sequence, or may have lost it permanently \citep[e.g.][]{Ramirez2014,Tian2015,Luger2015,Barnes2016}.

To account for this caveat, more detailed future studies could estimate and subtract the amount of time for which the planet was outside of the habitable zone from its age before performing a correlation test. In contrast, planets which have been rendered permanently uninhabitable by the pre-main sequence star represent a reduction in \flife, and are therefore already factored into our analysis.

\section{Conclusions}

Motivated by a new generation of space missions concepts that aim to search for atmospheric biosignatures across statistically meaningful samples of planets, we explore how constraints on the presence of atmospheric \Otwo\ or \Othree\ as a function of age could be used to study how inhabited planets and life co-evolve, and to test the robustness of oxygen as a biosignature. A possible starting hypothesis for the evolution of inhabited planets is that their atmospheres evolve in a similar manner to Earth's. We show that this null hypothesis predicts a strong, positive age-oxygen correlation among such worlds. The presence of such a trend, if detected by future observatories, could serve as additional evidence for life.

We show that by detecting or rejecting the presence of \Otwo\ -- or its byproduct \Othree\  -- in a sufficiently large sample of potentially habitable planets, it will be possible to confirm this hypothesis using statistical correlation tests -- without needing to know which planets are inhabited. To confidently detect the age-oxygen correlation, we find that a sample size of $\sim 20$ potentially habitable planets must be observed if $\sim 80\%$ of them are in fact inhabited. If the inhabited fraction is only $\sim 10\%$, then $\sim 300$ planets must be observed, and if it is smaller than $\sim$2\% then more than one thousand planets are required. These sample sizes are similar to those predicted for ambitious space missions proposed to launch within the coming decades. 

Our results have important implications for the target selection of future biosignature surveys which should be considered as their missions are designed and built. Namely, surveys which must down-select or prioritize their target lists should favor young and old over intermediate-age stars, as doing so could reduce the number of planets required to detect the age-oxygen correlation by about 35\%. Similarly, surveys should avoid excluding young stars from their target lists, as this could increase the number of planets required by a factor of $\sim 3$. Such target selection strategies will require precise stellar age measurements, which should be achievable for many FGK dwarfs through asteroseismology, but are yet out of reach for low-mass stars.

These results are also sensitive to the abundance of abiotically-produced \Otwo\ in exoplanet atmospheres. While Earth has no substantial non-biological oxygen sources, it has been proposed that a small fraction of potentially habitable exoplanets could. If this fraction exceeds $\sim 10\%$, then many more planets must be characterized to detect the age-oxygen correlation.

Our study offers a promising initial analysis of the capacity of next-generation observatories to study the oxygen evolution of habitable planets. Future studies can expand upon this groundwork by incorporating quantitative treatments of the influence of planet size, composition, stellar environment, and other factors which may impact the proposed age-oxygen correlation. Finally, while in-depth studies of individual planets -- especially those presenting biosignatures -- will be invaluable, statistical analyses could enable a broader understanding of life as a universal phenomenon. We encourage that this and other statistical hypotheses be given proper consideration as new telescopes and instruments are being designed to characterize the atmospheres of habitable exoplanets.

\acknowledgements
We thank Regis Ferriere, Stephane Mezevet, Mercedes L\'opez-Morales, and Nicolas Cowan for discussions which contributed to this work. A.B. acknowledges support from the NASA Earth and Space Science Fellowship Program under grant No. 80NSS\-C17K0470. The results reported herein benefited from collaborations and/or information exchange within NASA's Nexus for Exoplanet System Science (NExSS) research coordination network sponsored by NASA's Science Mission Directorate. This research has made use of NASA's Astrophysics Data System.

\software{Matplotlib \citep{Hunter07}, NumPy \citep{Oliphant06}, scikit-image \citep{VanDerWalt2014}, SciPy \citep{Jones01}}

\bibliography{references}{}

\begin{thebibliography}{}
\expandafter\ifx\csname natexlab\endcsname\relax\def\natexlab#1{#1}\fi
\providecommand{\url}[1]{\href{#1}{#1}}
\providecommand{\dodoi}[1]{doi:~\href{http://doi.org/#1}{\nolinkurl{#1}}}
\providecommand{\doeprint}[1]{\href{http://ascl.net/#1}{\nolinkurl{http://ascl.net/#1}}}
\providecommand{\doarXiv}[1]{\href{https://arxiv.org/abs/#1}{\nolinkurl{https://arxiv.org/abs/#1}}}

\bibitem[{{Aizawa} {et~al.}(2020){Aizawa}, {Kawahara}, \& {Fan}}]{Aizawa2020}
{Aizawa}, M., {Kawahara}, H., \& {Fan}, S. 2020, arXiv e-prints,
  arXiv:2004.03941.
\newblock \doarXiv{2004.03941}

\bibitem[{Anbar {et~al.}(2007)Anbar, Duan, Lyons, Arnold, Kendall, Creaser,
  Kaufman, Gordon, Scott, Garvin, \& Buick}]{Anbar2007}
Anbar, A.~D., Duan, Y., Lyons, T.~W., {et~al.} 2007, Science, 317, 1903,
  \dodoi{10.1126/science.1140325}

\bibitem[{Angel {et~al.}(1986)Angel, Cheng, \& Woolf}]{Angel1986}
Angel, J.~R., Cheng, A.~Y., \& Woolf, N.~J. 1986, Nature, 322, 341,
  \dodoi{10.1038/322341a0}

\bibitem[{Apai {et~al.}(2019)Apai, Milster, Kim, Bixel, Schneider, Liang, \&
  Arenberg}]{Apai2019}
Apai, D., Milster, T.~D., Kim, D.~W., {et~al.} 2019, The Astronomical Journal,
  158, 83, \dodoi{10.3847/1538-3881/ab2631}

\bibitem[{{Apai} {et~al.}(2018){Apai}, {Rackham}, {Giampapa}, {Angerhausen},
  {Teske}, {Barstow}, {Carone}, {Cegla}, {Domagal-Goldman}, {Espinoza},
  {Giles}, {Gully-Santiago}, {Haywood}, {Hu}, {Jordan}, {Kreidberg}, {Line},
  {Llama}, {L{\'o}pez-Morales}, {Marley}, \& {de Wit}}]{Apai2018}
{Apai}, D., {Rackham}, B.~V., {Giampapa}, M.~S., {et~al.} 2018, arXiv e-prints,
  arXiv:1803.08708.
\newblock \doarXiv{1803.08708}

\bibitem[{Arney {et~al.}(2018)Arney, Domagal-Goldman, \& Meadows}]{Arney2018}
Arney, G., Domagal-Goldman, S.~D., \& Meadows, V.~S. 2018, Astrobiology, 18,
  311, \dodoi{10.1089/ast.2017.1666}

\bibitem[{Arney {et~al.}(2016)Arney, Domagal-Goldman, Meadows, Wolf,
  Schwieterman, Charnay, Claire, H{\'{e}}brard, \& Trainer}]{Arney2016}
Arney, G., Domagal-Goldman, S.~D., Meadows, V.~S., {et~al.} 2016, Astrobiology,
  16, 873, \dodoi{10.1089/ast.2015.1422}

\bibitem[{Arney {et~al.}(2017)Arney, Meadows, Domagal-Goldman, Deming,
  Robinson, Tovar, Wolf, \& Schwieterman}]{Arney2017}
Arney, G.~N., Meadows, V.~S., Domagal-Goldman, S.~D., {et~al.} 2017, The
  Astrophysical Journal, 836, 49, \dodoi{10.3847/1538-4357/836/1/49}

\bibitem[{{Baran} {et~al.}(2011{\natexlab{a}}){Baran}, {Fox-Machado}, {Lykke},
  {Nielsen}, \& {Telting}}]{Baran2011b}
{Baran}, A.~S., {Fox-Machado}, L., {Lykke}, J., {Nielsen}, M., \& {Telting},
  J.~H. 2011{\natexlab{a}}, \actaa, 61, 325

\bibitem[{{Baran} {et~al.}(2013){Baran}, {Winiarski}, {Siwak}, {Fox-Machado},
  {Kozie{\l}-Wierzbowska}, {Krzesinski}, {Dr{\'o}{\.z}dz}, \&
  {Winans}}]{Baran2013}
{Baran}, A.~S., {Winiarski}, M., {Siwak}, M., {et~al.} 2013, \actaa, 63, 41

\bibitem[{{Baran} {et~al.}(2011{\natexlab{b}}){Baran}, {Winiarski},
  {Krzesi{\'n}ski}, {Fox-Machado}, {Kawaler}, {Dr{\'o}{\.z}dz}, {Faltenbacher},
  {Thompson}, \& {Reed}}]{Baran2011a}
{Baran}, A.~S., {Winiarski}, M., {Krzesi{\'n}ski}, J., {et~al.}
  2011{\natexlab{b}}, \actaa, 61, 37

\bibitem[{Barnes {et~al.}(2016)Barnes, Deitrick, Luger, Driscoll, Quinn,
  Fleming, Guyer, McDonald, Meadows, Arney, Crisp, Domagal-Goldman,
  Foreman-Mackey, Kaib, Lincowski, Lustig-Yaeger, \& Schwieterman}]{Barnes2016}
Barnes, R., Deitrick, R., Luger, R., {et~al.} 2016.
\newblock \doarXiv{1608.06919}

\bibitem[{{Bean} {et~al.}(2017){Bean}, {Abbot}, \& {Kempton}}]{Bean2017}
{Bean}, J.~L., {Abbot}, D.~S., \& {Kempton}, E. M.~R. 2017, \apjl, 841, L24,
  \dodoi{10.3847/2041-8213/aa738a}

\bibitem[{{Berdi{\~n}as} {et~al.}(2017){Berdi{\~n}as},
  {Rodr{\'\i}guez-L{\'o}pez}, {Amado}, {Anglada-Escud{\'e}}, {Barnes},
  {MacDonald}, {Zechmeister}, \& {Sarmiento}}]{Berdinas2017}
{Berdi{\~n}as}, Z.~M., {Rodr{\'\i}guez-L{\'o}pez}, C., {Amado}, P.~J., {et~al.}
  2017, \mnras, 469, 4268, \dodoi{10.1093/mnras/stx1140}

\bibitem[{Brocks {et~al.}(1999)Brocks, Logan, Buick, \& Summons}]{Brocks1999}
Brocks, J.~J., Logan, G.~A., Buick, R., \& Summons, R.~E. 1999, Science, 285,
  1033, \dodoi{10.1126/science.285.5430.1033}

\bibitem[{Buccino {et~al.}(2007)Buccino, Lemarchand, \& Mauas}]{Buccino2007}
Buccino, A.~P., Lemarchand, G.~A., \& Mauas, P.~J. 2007, Icarus, 192, 582,
  \dodoi{10.1016/j.icarus.2007.08.012}

\bibitem[{{Burgasser} \& {Mamajek}(2017)}]{Burgasser2017}
{Burgasser}, A.~J., \& {Mamajek}, E.~E. 2017, \apj, 845, 110,
  \dodoi{10.3847/1538-4357/aa7fea}

\bibitem[{Catling {et~al.}(2005)Catling, Glein, Zahnle, \& McKay}]{Catling2005}
Catling, D.~C., Glein, C.~R., Zahnle, K.~J., \& McKay, C.~P. 2005, {Why O2 is
  required by complex life on habitable planets and the concept of planetary
  "Oxygenation time"},  Mary Ann Liebert, Inc. 2 Madison Avenue Larchmont, NY
  10538 USA, \dodoi{10.1089/ast.2005.5.415}

\bibitem[{Catling \& Zahnle(2020)}]{Catling2020}
Catling, D.~C., \& Zahnle, K.~J. 2020, Science Advances, 6, eaax1420,
  \dodoi{10.1126/sciadv.aax1420}

\bibitem[{Catling {et~al.}(2001)Catling, Zahnle, \& McKay}]{Catling2001}
Catling, D.~C., Zahnle, K.~J., \& McKay, C.~P. 2001, Science, 293, 839,
  \dodoi{10.1126/science.1061976}

\bibitem[{{Chaplin} {et~al.}(2014){Chaplin}, {Basu}, {Huber}, {Serenelli},
  {Casagrande}, {Silva Aguirre}, {Ball}, {Creevey}, {Gizon}, {Handberg},
  {Karoff}, {Lutz}, {Marques}, {Miglio}, {Stello}, {Suran}, {Pricopi},
  {Metcalfe}, {Monteiro}, {Molenda-{\.Z}akowicz}, {Appourchaux},
  {Christensen-Dalsgaard}, {Elsworth}, {Garc{\'\i}a}, {Houdek}, {Kjeldsen},
  {Bonanno}, {Campante}, {Corsaro}, {Gaulme}, {Hekker}, {Mathur}, {Mosser},
  {R{\'e}gulo}, \& {Salabert}}]{Chaplin2014}
{Chaplin}, W.~J., {Basu}, S., {Huber}, D., {et~al.} 2014, \apjs, 210, 1,
  \dodoi{10.1088/0067-0049/210/1/1}

\bibitem[{{Checlair} {et~al.}(2019){Checlair}, {Abbot}, {Webber}, {Feng},
  {Bean}, {Schwieterman}, {Stark}, {Robinson}, {Kempton}, {Alcabes}, {Apai},
  {Arney}, {Cowan}, {Domagal-Goldman}, {Dong}, {Fleming}, {Fujii}, {Graham},
  {Guzewich}, {Hasegawa}, {Hayworth}, {Kane}, {Kite}, {Komacek}, {Kopparapu},
  {Mansfield}, {Marounina}, {Montet}, {Olson}, {Paradise}, {Popovic},
  {Rackham}, {Ramirez}, {Rau}, {Reinhard}, {Renaud}, {Rogers}, {Walkowicz},
  {Warren}, \& {Wolf}}]{Checlair2019}
{Checlair}, J., {Abbot}, D.~S., {Webber}, R.~J., {et~al.} 2019, \baas, 51, 404.
\newblock \doarXiv{1903.05211}

\bibitem[{Claire {et~al.}(2006)Claire, Catling, \& Zahnle}]{Claire2006}
Claire, M.~W., Catling, D.~C., \& Zahnle, K.~J. 2006, Geobiology, 4, 239,
  \dodoi{10.1111/j.1472-4669.2006.00084.x}

\bibitem[{Cowan \& Fujii(2018)}]{Cowan2018}
Cowan, N.~B., \& Fujii, Y. 2018, in Handbook of Exoplanets (Cham: Springer
  International Publishing), 1469--1484, \dodoi{10.1007/978-3-319-55333-7_147}

\bibitem[{Cowan {et~al.}(2009)Cowan, Agol, Meadows, Robinson, Livengood,
  Deming, Lisse, A'Hearn, Wellnitz, Seager, Charbonneau, \& Team}]{Cowan2009}
Cowan, N.~B., Agol, E., Meadows, V.~S., {et~al.} 2009, The Astrophysical
  Journal, 700, 915, \dodoi{10.1088/0004-637X/700/2/915}

\bibitem[{{Creevey} {et~al.}(2017){Creevey}, {Metcalfe}, {Schultheis},
  {Salabert}, {Bazot}, {Th{\'e}venin}, {Mathur}, {Xu}, \&
  {Garc{\'\i}a}}]{Creevey2017}
{Creevey}, O.~L., {Metcalfe}, T.~S., {Schultheis}, M., {et~al.} 2017, \aap,
  601, A67, \dodoi{10.1051/0004-6361/201629496}

\bibitem[{{Des Marais} {et~al.}(2002){Des Marais}, Harwit, Jucks, Kasting, Lin,
  Lunine, Schneider, Seager, Traub, \& Woolf}]{DesMarais2002}
{Des Marais}, D.~J., Harwit, M.~O., Jucks, K.~W., {et~al.} 2002, Astrobiology,
  2, 153, \dodoi{10.1089/15311070260192246}

\bibitem[{Domagal-Goldman {et~al.}(2011)Domagal-Goldman, Meadows, Claire, \&
  Kasting}]{Domagal-Goldman2011}
Domagal-Goldman, S.~D., Meadows, V.~S., Claire, M.~W., \& Kasting, J.~F. 2011,
  Astrobiology, 11, 419, \dodoi{10.1089/ast.2010.0509}

\bibitem[{Domagal-Goldman {et~al.}(2014)Domagal-Goldman, Segura, Claire,
  Robinson, \& Meadows}]{Domagal-Goldman2014}
Domagal-Goldman, S.~D., Segura, A., Claire, M.~W., Robinson, T.~D., \& Meadows,
  V.~S. 2014, Astrophysical Journal, 792, \dodoi{10.1088/0004-637X/792/2/90}

\bibitem[{Driscoll \& Barnes(2015)}]{Driscoll2015}
Driscoll, P.~E., \& Barnes, R. 2015, Astrobiology, 15, 739,
  \dodoi{10.1089/ast.2015.1325}

\bibitem[{Fan {et~al.}(2019)Fan, Li, Li, Bartlett, Jiang, Natraj, Crisp, \&
  Yung}]{Fan2019}
Fan, S., Li, C., Li, J.-Z., {et~al.} 2019, The Astrophysical Journal, 882, L1,
  \dodoi{10.3847/2041-8213/ab3a49}

\bibitem[{{Fantin} {et~al.}(2019){Fantin}, {C{\^o}t{\'e}}, {McConnachie},
  {Bergeron}, {Cuilland re}, {Gwyn}, {Ibata}, {Thomas}, {Carlberg}, {Fabbro},
  {Haywood}, {Lan{\c{c}}on}, {Lewis}, {Malhan}, {Martin}, {Navarro}, {Scott},
  \& {Starkenburg}}]{Fantin2019}
{Fantin}, N.~J., {C{\^o}t{\'e}}, P., {McConnachie}, A.~W., {et~al.} 2019, \apj,
  887, 148, \dodoi{10.3847/1538-4357/ab5521}

\bibitem[{Farr {et~al.}(2018)Farr, Farr, Cowan, Haggard, \&
  Robinson}]{Farr2018}
Farr, B., Farr, W.~M., Cowan, N.~B., Haggard, H.~M., \& Robinson, T. 2018, The
  Astronomical Journal, 156, 146, \dodoi{10.3847/1538-3881/aad775}

\bibitem[{{Ford} {et~al.}(2001){Ford}, {Seager}, \& {Turner}}]{Ford2001}
{Ford}, E.~B., {Seager}, S., \& {Turner}, E.~L. 2001, \nat, 412, 885,
  \dodoi{10.1038/35091009}

\bibitem[{Gale \& Wandel(2016)}]{Gale2016}
Gale, J., \& Wandel, A. 2016, International Journal of Astrobiology, 16, 1,
  \dodoi{10.1017/S1473550415000440}

\bibitem[{Gao {et~al.}(2015)Gao, Hu, Robinson, Li, \& Yung}]{Gao2015}
Gao, P., Hu, R., Robinson, T.~D., Li, C., \& Yung, Y.~L. 2015, Astrophysical
  Journal, 806, \dodoi{10.1088/0004-637X/806/2/249}

\bibitem[{Gillon {et~al.}(2017)Gillon, Triaud, Demory, Jehin, Agol, Deck,
  Lederer, de~Wit, Burdanov, Ingalls, Bolmont, Leconte, Raymond, Selsis,
  Turbet, Barkaoui, Burgasser, Burleigh, Carey, Chaushev, Copperwheat, Delrez,
  Fernandes, Holdsworth, Kotze, {Van Grootel}, Almleaky, Benkhaldoun, Magain,
  \& Queloz}]{Gillon2017}
Gillon, M., Triaud, A. H. M.~J., Demory, B.-O., {et~al.} 2017, Nature, 542,
  456, \dodoi{10.1038/nature21360}

\bibitem[{Haqq-Misra(2019)}]{Haqq-Misra2019}
Haqq-Misra, J. 2019, Astrobiology, 19, 1292, \dodoi{10.1089/ast.2018.1946}

\bibitem[{Haqq-Misra \& Kopparapu(2018)}]{Haqq-Misra2018}
Haqq-Misra, J., \& Kopparapu, R.~K. 2018, in Habitability of the Universe
  Before Earth (Elsevier), 307--319, \dodoi{10.1016/b978-0-12-811940-2.00013-7}

\bibitem[{Harman {et~al.}(2018)Harman, Felton, Hu, Domagal-Goldman, Segura,
  Tian, \& Kasting}]{Harman2018}
Harman, C.~E., Felton, R., Hu, R., {et~al.} 2018, The Astrophysical Journal,
  866, 56, \dodoi{10.3847/1538-4357/aadd9b}

\bibitem[{Harman {et~al.}(2015)Harman, Schwieterman, Schottelkotte, \&
  Kasting}]{Harman2015}
Harman, C.~E., Schwieterman, E.~W., Schottelkotte, J.~C., \& Kasting, J.~F.
  2015, Astrophysical Journal, 812, \dodoi{10.1088/0004-637X/812/2/137}

\bibitem[{{Hu} {et~al.}(2020){Hu}, {Peterson}, \& {Wolf}}]{Hu2020}
{Hu}, R., {Peterson}, L., \& {Wolf}, E.~T. 2020, \apj, 888, 122,
  \dodoi{10.3847/1538-4357/ab5f07}

\bibitem[{Hu {et~al.}(2012)Hu, Seager, \& Bains}]{Hu2012}
Hu, R., Seager, S., \& Bains, W. 2012, Astrophysical Journal, 761,
  \dodoi{10.1088/0004-637X/761/2/166}

\bibitem[{Hunten \& Donahue(1976)}]{Hunten1976}
Hunten, D.~M., \& Donahue, T.~M. 1976, Annual Review of Earth and Planetary
  Sciences, 4, 265, \dodoi{10.1146/annurev.ea.04.050176.001405}

\bibitem[{Hunter(2007)}]{Hunter07}
Hunter, J.~D. 2007, Computing In Science \& Engineering, 9, 90,
  \dodoi{10.1109/MCSE.2007.55}

\bibitem[{{Iyer} \& {Line}(2019)}]{IyerLine2019}
{Iyer}, A.~R., \& {Line}, M.~R. 2019, arXiv e-prints, arXiv:1912.04389.
\newblock \doarXiv{1912.04389}

\bibitem[{James \& Hu(2018)}]{James2018}
James, T., \& Hu, R. 2018, The Astrophysical Journal, 867, 17,
  \dodoi{10.3847/1538-4357/aae2bb}

\bibitem[{Jones {et~al.}(2001)Jones, Oliphant, Peterson, {et~al.}}]{Jones01}
Jones, E., Oliphant, T., Peterson, P., {et~al.} 2001, {SciPy}: Open source
  scientific tools for {Python}.
\newblock \url{http://www.scipy.org/}

\bibitem[{{Kammerer} \& {Quanz}(2018)}]{Kammerer2018}
{Kammerer}, J., \& {Quanz}, S.~P. 2018, \aap, 609, A4,
  \dodoi{10.1051/0004-6361/201731254}

\bibitem[{Kasting {et~al.}(1993)Kasting, Whitmire, \& Reynolds}]{Kasting1993}
Kasting, J.~F., Whitmire, D.~P., \& Reynolds, R.~T. 1993, Icarus, 101, 108,
  \dodoi{10.1006/icar.1993.1010}

\bibitem[{{Kawashima} \& {Rugheimer}(2019)}]{Kawashima2019}
{Kawashima}, Y., \& {Rugheimer}, S. 2019, \aj, 157, 213,
  \dodoi{10.3847/1538-3881/ab14e3}

\bibitem[{{Kayhan} {et~al.}(2019){Kayhan}, {Y{\i}ld{\i}z}, \& {{\c{C}}elik
  Orhan}}]{Kayhan2019}
{Kayhan}, C., {Y{\i}ld{\i}z}, M., \& {{\c{C}}elik Orhan}, Z. 2019, \mnras, 490,
  1509, \dodoi{10.1093/mnras/stz2634}

\bibitem[{Kendall {et~al.}(2010)Kendall, Reinhard, Lyons, Kaufman, Poulton, \&
  Anbar}]{Kendall2010}
Kendall, B., Reinhard, C.~T., Lyons, T.~W., {et~al.} 2010, Nature Geoscience,
  3, 647, \dodoi{10.1038/ngeo942}

\bibitem[{Kiang {et~al.}(2007)Kiang, Segura, Tinetti, Govindjee, Blankenship,
  Cohen, Siefert, Crisp, \& Meadows}]{Kiang2007}
Kiang, N.~Y., Segura, A., Tinetti, G., {et~al.} 2007, Astrobiology, 7, 252,
  \dodoi{10.1089/ast.2006.0108}

\bibitem[{Kopparapu {et~al.}(2014)Kopparapu, Ramirez, Schottelkotte, Kasting,
  Domagal-Goldman, \& Eymet}]{Kopparapu2014}
Kopparapu, R.~K., Ramirez, R.~M., Schottelkotte, J., {et~al.} 2014,
  Astrophysical Journal Letters, 787, L29, \dodoi{10.1088/2041-8205/787/2/L29}

\bibitem[{Kopparapu {et~al.}(2013)Kopparapu, Ramirez, Kasting, Eymet, Robinson,
  Mahadevan, Terrien, Domagal-Goldman, Meadows, \& Deshpande}]{Kopparapu2013}
Kopparapu, R.~K., Ramirez, R., Kasting, J.~F., {et~al.} 2013, Astrophysical
  Journal, 765, 131, \dodoi{10.1088/0004-637X/765/2/131}

\bibitem[{Kopparapu {et~al.}(2018)Kopparapu, H{\'{e}}brard, Belikov, Batalha,
  Mulders, Stark, Teal, Domagal-Goldman, \& Mandell}]{Kopparapu2018}
Kopparapu, R.~K., H{\'{e}}brard, E., Belikov, R., {et~al.} 2018, The
  Astrophysical Journal, 856, 122, \dodoi{10.3847/1538-4357/aab205}

\bibitem[{Krissansen-Totton {et~al.}(2018)Krissansen-Totton, Olson, \&
  Catling}]{Krissansen-Totton2018}
Krissansen-Totton, J., Olson, S., \& Catling, D.~C. 2018, Science Advances, 4,
  \dodoi{10.1126/sciadv.aao5747}

\bibitem[{{Krzesinski} {et~al.}(2012){Krzesinski}, {Baran}, {Winiarski},
  {Fox-Machado}, {Dr{\'o}{\.z}d{\.z}}, {Siwak}, \&
  {Kozie{\l}-Wierzbowska}}]{Krzesinski2012}
{Krzesinski}, J., {Baran}, A.~S., {Winiarski}, M., {et~al.} 2012, \actaa, 62,
  201

\bibitem[{Kurzweil {et~al.}(2013)Kurzweil, Claire, Thomazo, Peters, Hannington,
  \& Strauss}]{Kurzweil2013}
Kurzweil, F., Claire, M., Thomazo, C., {et~al.} 2013, Earth and Planetary
  Science Letters, 366, 17, \dodoi{10.1016/j.epsl.2013.01.028}

\bibitem[{Lehmer {et~al.}(2018)Lehmer, Catling, Parenteau, \&
  Hoehler}]{Lehmer2018}
Lehmer, O.~R., Catling, D.~C., Parenteau, M.~N., \& Hoehler, T.~M. 2018, The
  Astrophysical Journal, 859, 171, \dodoi{10.3847/1538-4357/aac104}

\bibitem[{Lingam \& Loeb(2018)}]{Lingam2018}
Lingam, M., \& Loeb, A. 2018, Reviews of Modern Physics, 91, 021002,
  \dodoi{10.1103/RevModPhys.91.021002}

\bibitem[{Lingam \& Loeb(2019)}]{Lingam2019}
---. 2019, Monthly Notices of the Royal Astronomical Society, 485, 5924,
  \dodoi{10.1093/mnras/stz847}

\bibitem[{Lovelock(1965)}]{Lovelock1965}
Lovelock, J.~E. 1965, Nature, 207, 568, \dodoi{10.1038/207568a0}

\bibitem[{Luger \& Barnes(2015)}]{Luger2015}
Luger, R., \& Barnes, R. 2015, Astrobiology, 15, 119,
  \dodoi{10.1089/ast.2014.1231}

\bibitem[{{Lund} {et~al.}(2019){Lund}, {Knudstrup}, {Silva Aguirre}, {Basu},
  {Chontos}, {Von Essen}, {Chaplin}, {Bieryla}, {Casagrande}, {Vanderburg},
  {Huber}, {Kane}, {Albrecht}, {Latham}, {Davies}, {Becker}, \&
  {Rodriguez}}]{Lund2019}
{Lund}, M.~N., {Knudstrup}, E., {Silva Aguirre}, V., {et~al.} 2019, \aj, 158,
  248, \dodoi{10.3847/1538-3881/ab5280}

\bibitem[{Lunine(2013)}]{Lunine2013}
Lunine, J.~I. 2013, {Earth : evolution of a habitable world} (Cambridge
  University Press), 318

\bibitem[{{Lustig-Yaeger} {et~al.}(2019){Lustig-Yaeger}, {Meadows}, \&
  {Lincowski}}]{Lustig-Yaeger2019}
{Lustig-Yaeger}, J., {Meadows}, V.~S., \& {Lincowski}, A.~P. 2019, \aj, 158,
  27, \dodoi{10.3847/1538-3881/ab21e0}

\bibitem[{Lustig-Yaeger {et~al.}(2018)Lustig-Yaeger, Meadows, {Tovar Mendoza},
  Schwieterman, Fujii, Luger, \& Robinson}]{Lustig-Yaeger2018}
Lustig-Yaeger, J., Meadows, V.~S., {Tovar Mendoza}, G., {et~al.} 2018, The
  Astronomical Journal, 156, 301, \dodoi{10.3847/1538-3881/aaed3a}

\bibitem[{Lyons {et~al.}(2014)Lyons, Reinhard, \& Planavsky}]{Lyons2014}
Lyons, T.~W., Reinhard, C.~T., \& Planavsky, N.~J. 2014, {The rise of oxygen in
  Earth's early ocean and atmosphere}, \dodoi{10.1038/nature13068}

\bibitem[{Mann \& Whitney(1947)}]{Mann1947}
Mann, H.~B., \& Whitney, D.~R. 1947, The Annals of Mathematical Statistics, 18,
  50, \dodoi{10.1214/aoms/1177730491}

\bibitem[{{Mathur} {et~al.}(2012){Mathur}, {Metcalfe}, {Woitaszek}, {Bruntt},
  {Verner}, {Christensen-Dalsgaard}, {Creevey}, {Do{\v{g}}an}, {Basu},
  {Karoff}, {Stello}, {Appourchaux}, {Campante}, {Chaplin}, {Garc{\'\i}a},
  {Bedding}, {Benomar}, {Bonanno}, {Deheuvels}, {Elsworth}, {Gaulme}, {Guzik},
  {Handberg}, {Hekker}, {Herzberg}, {Monteiro}, {Piau}, {Quirion},
  {R{\'e}gulo}, {Roth}, {Salabert}, {Serenelli}, {Thompson}, {Trampedach},
  {White}, {Ballot}, {Brand{\~a}o}, {Molenda-{\.Z}akowicz}, {Kjeldsen},
  {Twicken}, {Uddin}, \& {Wohler}}]{Mathur2012}
{Mathur}, S., {Metcalfe}, T.~S., {Woitaszek}, M., {et~al.} 2012, \apj, 749,
  152, \dodoi{10.1088/0004-637X/749/2/152}

\bibitem[{Meadows(2017)}]{Meadows2017}
Meadows, V.~S. 2017, {Reflections on O2 as a Biosignature in Exoplanetary
  Atmospheres},  Mary Ann Liebert Inc., \dodoi{10.1089/ast.2016.1578}

\bibitem[{Meadows {et~al.}(2018{\natexlab{a}})Meadows, Reinhard, Arney,
  Parenteau, Schwieterman, Domagal-Goldman, Lincowski, Stapelfeldt, Rauer,
  DasSarma, Hegde, Narita, Deitrick, Lustig-Yaeger, Lyons, Siegler, \&
  Grenfell}]{Meadows2018b}
Meadows, V.~S., Reinhard, C.~T., Arney, G.~N., {et~al.} 2018{\natexlab{a}},
  Astrobiology, 18, 630, \dodoi{10.1089/ast.2017.1727}

\bibitem[{Meadows {et~al.}(2018{\natexlab{b}})Meadows, Arney, Schwieterman,
  Lustig-Yaeger, Lincowski, Robinson, Domagal-Goldman, Deitrick, Barnes,
  Fleming, Luger, Driscoll, Quinn, \& Crisp}]{Meadows2018a}
Meadows, V.~S., Arney, G.~N., Schwieterman, E.~W., {et~al.} 2018{\natexlab{b}},
  Astrobiology, 18, 133, \dodoi{10.1089/ast.2016.1589}

\bibitem[{{Mor} {et~al.}(2019){Mor}, {Robin}, {Figueras}, {Roca-F{\`a}brega},
  \& {Luri}}]{Mor2019}
{Mor}, R., {Robin}, A.~C., {Figueras}, F., {Roca-F{\`a}brega}, S., \& {Luri},
  X. 2019, \aap, 624, L1, \dodoi{10.1051/0004-6361/201935105}

\bibitem[{{Mullan} \& {Bais}(2018)}]{Mullan2018}
{Mullan}, D.~J., \& {Bais}, H.~P. 2018, \apj, 865, 101,
  \dodoi{10.3847/1538-4357/aadfd1}

\bibitem[{Narita {et~al.}(2015)Narita, Enomoto, Masaoka, \&
  Kusakabe}]{Narita2015}
Narita, N., Enomoto, T., Masaoka, S., \& Kusakabe, N. 2015, Scientific Reports,
  5, \dodoi{10.1038/srep13977}

\bibitem[{Neil \& Rogers(2020)}]{Neil2020}
Neil, A.~R., \& Rogers, L.~A. 2020, The Astrophysical Journal, 891, 12,
  \dodoi{10.3847/1538-4357/ab6a92}

\bibitem[{Oliphant(2006)}]{Oliphant06}
Oliphant, T.~E. 2006, Guide to NumPy, Provo, UT.
\newblock \url{http://www.tramy.us/}

\bibitem[{{Origins Space Telescope Study Team}(2019)}]{Origins2019}
{Origins Space Telescope Study Team}. 2019, {Origins Space Telescope Mission
  Concept Study Report}, Tech. rep.

\bibitem[{Owen(1980)}]{Owen1980}
Owen, T. 1980 (Springer, Dordrecht), 177--185,
  \dodoi{10.1007/978-94-009-9115-6_17}

\bibitem[{{Pascucci} {et~al.}(2019){Pascucci}, {Mulders}, \&
  {Lopez}}]{Pascucci2019}
{Pascucci}, I., {Mulders}, G.~D., \& {Lopez}, E. 2019, \apjl, 883, L15,
  \dodoi{10.3847/2041-8213/ab3dac}

\bibitem[{Pilcher(2003)}]{Pilcher2003}
Pilcher, C.~B. 2003in  (Mary Ann Liebert, Inc.), 471--486,
  \dodoi{10.1089/153110703322610582}

\bibitem[{{Pinhas} {et~al.}(2018){Pinhas}, {Rackham}, {Madhusudhan}, \&
  {Apai}}]{Pinhas2018}
{Pinhas}, A., {Rackham}, B.~V., {Madhusudhan}, N., \& {Apai}, D. 2018, \mnras,
  480, 5314, \dodoi{10.1093/mnras/sty2209}

\bibitem[{Planavsky {et~al.}(2014)Planavsky, Asael, Hofmann, Reinhard, Lalonde,
  Knudsen, Wang, {Ossa Ossa}, Pecoits, Smith, Beukes, Bekker, Johnson,
  Konhauser, Lyons, \& Rouxel}]{Planavsky2014}
Planavsky, N.~J., Asael, D., Hofmann, A., {et~al.} 2014, Nature Geoscience, 7,
  283, \dodoi{10.1038/ngeo2122}

\bibitem[{{Quanz} {et~al.}(2018){Quanz}, {Kammerer}, {Defr{\`e}re}, {Absil},
  {Glauser}, \& {Kitzmann}}]{Quanz2018}
{Quanz}, S.~P., {Kammerer}, J., {Defr{\`e}re}, D., {et~al.} 2018, in Society of
  Photo-Optical Instrumentation Engineers (SPIE) Conference Series, Vol. 10701,
  \procspie, 107011I, \dodoi{10.1117/12.2312051}

\bibitem[{{Rackham} {et~al.}(2019{\natexlab{a}}){Rackham}, {Pinhas}, {Apai},
  {Haywood}, {Cegla}, {Espinoza}, {Teske}, {Gully-Santiago}, {Rau}, {Morris},
  {Angerhausen}, {Barclay}, {Carone}, {Cauley}, {de Wit}, {Domagal-Goldman},
  {Dong}, {Dragomir}, {Giampapa}, {Hasegawa}, {Hinkel}, {Hu}, {Jord{\'a}n},
  {Kitiashvili}, {Kreidberg}, {Lisse}, {Llama}, {L{\'o}pez-Morales},
  {Mennesson}, {Molaverdikhani}, {Osip}, \& {Quintana}}]{Rackham2019b}
{Rackham}, B., {Pinhas}, A., {Apai}, D., {et~al.} 2019{\natexlab{a}}, \baas,
  51, 328.
\newblock \doarXiv{1903.06152}

\bibitem[{{Rackham} {et~al.}(2018){Rackham}, {Apai}, \&
  {Giampapa}}]{Rackham2018}
{Rackham}, B.~V., {Apai}, D., \& {Giampapa}, M.~S. 2018, \apj, 853, 122,
  \dodoi{10.3847/1538-4357/aaa08c}

\bibitem[{{Rackham} {et~al.}(2019{\natexlab{b}}){Rackham}, {Apai}, \&
  {Giampapa}}]{Rackham2019a}
---. 2019{\natexlab{b}}, \aj, 157, 96, \dodoi{10.3847/1538-3881/aaf892}

\bibitem[{{Ramirez} \& {Kaltenegger}(2014)}]{Ramirez2014}
{Ramirez}, R.~M., \& {Kaltenegger}, L. 2014, \apjl, 797, L25,
  \dodoi{10.1088/2041-8205/797/2/L25}

\bibitem[{{Ranjan} {et~al.}(2017){Ranjan}, {Wordsworth}, \&
  {Sasselov}}]{Ranjan2017}
{Ranjan}, S., {Wordsworth}, R., \& {Sasselov}, D.~D. 2017, \apj, 843, 110,
  \dodoi{10.3847/1538-4357/aa773e}

\bibitem[{{Rauer} {et~al.}(2014){Rauer}, {Catala}, {Aerts}, {Appourchaux},
  {Benz}, {Brandeker}, {Christensen-Dalsgaard}, {Deleuil}, {Gizon}, {Goupil},
  {G{\"u}del}, {Janot-Pacheco}, {Mas-Hesse}, {Pagano}, {Piotto}, {Pollacco},
  {Santos}, {Smith}, {Su{\'a}rez}, {Szab{\'o}}, {Udry}, {Adibekyan}, {Alibert},
  {Almenara}, {Amaro-Seoane}, {Eiff}, {Asplund}, {Antonello}, {Barnes},
  {Baudin}, {Belkacem}, {Bergemann}, {Bihain}, {Birch}, {Bonfils}, {Boisse},
  {Bonomo}, {Borsa}, {Brand {\~a}o}, {Brocato}, {Brun}, {Burleigh}, {Burston},
  {Cabrera}, {Cassisi}, {Chaplin}, {Charpinet}, {Chiappini}, {Church},
  {Csizmadia}, {Cunha}, {Damasso}, {Davies}, {Deeg}, {D{\'\i}az}, {Dreizler},
  {Dreyer}, {Eggenberger}, {Ehrenreich}, {Eigm{\"u}ller}, {Erikson}, {Farmer},
  {Feltzing}, {de Oliveira Fialho}, {Figueira}, {Forveille}, {Fridlund},
  {Garc{\'\i}a}, {Giommi}, {Giuffrida}, {Godolt}, {Gomes da Silva}, {Granzer},
  {Grenfell}, {Grotsch-Noels}, {G{\"u}nther}, {Haswell}, {Hatzes},
  {H{\'e}brard}, {Hekker}, {Helled}, {Heng}, {Jenkins}, {Johansen},
  {Khodachenko}, {Kislyakova}, {Kley}, {Kolb}, {Krivova}, {Kupka}, {Lammer},
  {Lanza}, {Lebreton}, {Magrin}, {Marcos-Arenal}, {Marrese}, {Marques},
  {Martins}, {Mathis}, {Mathur}, {Messina}, {Miglio}, {Montalban}, {Montalto},
  {Monteiro}, {Moradi}, {Moravveji}, {Mordasini}, {Morel}, {Mortier},
  {Nascimbeni}, {Nelson}, {Nielsen}, {Noack}, {Norton}, {Ofir}, {Oshagh},
  {Ouazzani}, {P{\'a}pics}, {Parro}, {Petit}, {Plez}, {Poretti}, {Quirrenbach},
  {Ragazzoni}, {Raimondo}, {Rainer}, {Reese}, {Redmer}, {Reffert},
  {Rojas-Ayala}, {Roxburgh}, {Salmon}, {Santerne}, {Schneider}, {Schou},
  {Schuh}, {Schunker}, {Silva-Valio}, {Silvotti}, {Skillen}, {Snellen}, {Sohl},
  {Sousa}, {Sozzetti}, {Stello}, {Strassmeier}, {{\v{S}}vanda}, {Szab{\'o}},
  {Tkachenko}, {Valencia}, {Van Grootel}, {Vauclair}, {Ventura}, {Wagner},
  {Walton}, {Weingrill}, {Werner}, {Wheatley}, \& {Zwintz}}]{Rauer2014}
{Rauer}, H., {Catala}, C., {Aerts}, C., {et~al.} 2014, Experimental Astronomy,
  38, 249, \dodoi{10.1007/s10686-014-9383-4}

\bibitem[{Reinhard {et~al.}(2017)Reinhard, Olson, Schwieterman, \&
  Lyons}]{Reinhard2017}
Reinhard, C.~T., Olson, S.~L., Schwieterman, E.~W., \& Lyons, T.~W. 2017,
  Astrobiology, 17, 287, \dodoi{10.1089/AST.2016.1598}

\bibitem[{Rimmer {et~al.}(2018)Rimmer, Xu, Thompson, Gillen, Sutherland, \&
  Queloz}]{Rimmer2018}
Rimmer, P.~B., Xu, J., Thompson, S.~J., {et~al.} 2018, Science Advances, 4,
  eaar3302, \dodoi{10.1126/sciadv.aar3302}

\bibitem[{Ritchie {et~al.}(2018)Ritchie, Larkum, \& Ribas}]{Ritchie2018}
Ritchie, R.~J., Larkum, A.~W., \& Ribas, I. 2018, International Journal of
  Astrobiology, 17, 147, \dodoi{10.1017/S1473550417000167}

\bibitem[{{Rodr{\'\i}guez} {et~al.}(2016){Rodr{\'\i}guez},
  {Rodr{\'\i}guez-L{\'o}pez}, {L{\'o}pez-Gonz{\'a}lez}, {Amado}, {Ocando}, \&
  {Berdi{\~n}as}}]{Rodriguez2016}
{Rodr{\'\i}guez}, E., {Rodr{\'\i}guez-L{\'o}pez}, C., {L{\'o}pez-Gonz{\'a}lez},
  M.~J., {et~al.} 2016, \mnras, 457, 1851, \dodoi{10.1093/mnras/stw033}

\bibitem[{{Rodr{\'\i}guez-L{\'o}pez} {et~al.}(2015){Rodr{\'\i}guez-L{\'o}pez},
  {Gizis}, {MacDonald}, {Amado}, \& {Carosso}}]{RodriguezLopez2015}
{Rodr{\'\i}guez-L{\'o}pez}, C., {Gizis}, J.~E., {MacDonald}, J., {Amado},
  P.~J., \& {Carosso}, A. 2015, \mnras, 446, 2613,
  \dodoi{10.1093/mnras/stu2211}

\bibitem[{{Rugheimer} \& {Kaltenegger}(2018)}]{Rugheimer2018}
{Rugheimer}, S., \& {Kaltenegger}, L. 2018, \apj, 854, 19,
  \dodoi{10.3847/1538-4357/aaa47a}

\bibitem[{{Schwieterman} {et~al.}(2018){Schwieterman}, {Kiang}, {Parenteau},
  {Harman}, {DasSarma}, {Fisher}, {Arney}, {Hartnett}, {Reinhard}, {Olson},
  {Meadows}, {Cockell}, {Walker}, {Grenfell}, {Hegde}, {Rugheimer}, {Hu}, \&
  {Lyons}}]{Schwieterman2018}
{Schwieterman}, E.~W., {Kiang}, N.~Y., {Parenteau}, M.~N., {et~al.} 2018,
  Astrobiology, 18, 663, \dodoi{10.1089/ast.2017.1729}

\bibitem[{Seager {et~al.}(2013)Seager, Bains, \& Hu}]{Seager2013}
Seager, S., Bains, W., \& Hu, R. 2013, Astrophysical Journal, 777, 95,
  \dodoi{10.1088/0004-637X/777/2/95}

\bibitem[{Segura {et~al.}(2003)Segura, Krelove, Kasting, Sommerlatt, Meadows,
  Crisp, Cohen, \& Mlawer}]{Segura2003}
Segura, A., Krelove, K., Kasting, J.~F., {et~al.} 2003, Astrobiology, 3, 689,
  \dodoi{10.1089/153110703322736024}

\bibitem[{{Silva Aguirre} {et~al.}(2015){Silva Aguirre}, {Davies}, {Basu},
  {Christensen-Dalsgaard}, {Creevey}, {Metcalfe}, {Bedding}, {Casagrande},
  {Handberg}, {Lund}, {Nissen}, {Chaplin}, {Huber}, {Serenelli}, {Stello}, {Van
  Eylen}, {Campante}, {Elsworth}, {Gilliland}, {Hekker}, {Karoff}, {Kawaler},
  {Kjeldsen}, \& {Lundkvist}}]{Silva2015}
{Silva Aguirre}, V., {Davies}, G.~R., {Basu}, S., {et~al.} 2015, \mnras, 452,
  2127, \dodoi{10.1093/mnras/stv1388}

\bibitem[{{Snaith} {et~al.}(2015){Snaith}, {Haywood}, {Di Matteo}, {Lehnert},
  {Combes}, {Katz}, \& {G{\'o}mez}}]{Snaith2015}
{Snaith}, O., {Haywood}, M., {Di Matteo}, P., {et~al.} 2015, \aap, 578, A87,
  \dodoi{10.1051/0004-6361/201424281}

\bibitem[{{Staguhn} {et~al.}(2019){Staguhn}, {Mandell}, {Stevenson}, {Saxena},
  {Kopparapu}, {Fixsen}, {Sharp}, {DiPirro}, {Knez}, {Wolf}, {Sotzen}, {Mandt},
  {Gong}, \& {Villanueva}}]{Staguhn2019}
{Staguhn}, J., {Mandell}, A., {Stevenson}, K., {et~al.} 2019, arXiv e-prints,
  arXiv:1908.02356.
\newblock \doarXiv{1908.02356}

\bibitem[{Stark {et~al.}(2015)Stark, Roberge, Mandell, Clampin,
  Domagal-Goldman, McElwain, \& Stapelfeldt}]{Stark2015}
Stark, C.~C., Roberge, A., Mandell, A., {et~al.} 2015, Astrophysical Journal,
  808, 149, \dodoi{10.1088/0004-637X/808/2/149}

\bibitem[{Stark {et~al.}(2014)Stark, Roberge, Mandell, \& Robinson}]{Stark2014}
Stark, C.~C., Roberge, A., Mandell, A., \& Robinson, T.~D. 2014, Astrophysical
  Journal, 795, 122, \dodoi{10.1088/0004-637X/795/2/122}

\bibitem[{Stark {et~al.}(2016{\natexlab{a}})Stark, Shaklan, Lisman, Cady,
  Savransky, Roberge, \& Mandell}]{Stark2016}
Stark, C.~C., Shaklan, S., Lisman, D., {et~al.} 2016{\natexlab{a}}, Journal of
  Astronomical Telescopes, Instruments, and Systems, 2, 041204,
  \dodoi{10.1117/1.JATIS.2.4.041204}

\bibitem[{Stark {et~al.}(2016{\natexlab{b}})Stark, Shaklan, Lisman, Cady,
  Savransky, Roberge, \& Mandell}]{Stark2016b}
---. 2016{\natexlab{b}}, Journal of Astronomical Telescopes, Instruments, and
  Systems, 2, 041204, \dodoi{10.1117/1.JATIS.2.4.041204}

\bibitem[{Stark {et~al.}(2019)Stark, Belikov, Bolcar, {Cady}, {Crill}, {Ertel},
  {Groff}, {Hildebrandt}, {Krist}, {Lisman}, {Mazoyer}, {Mennesson}, {Nemati},
  {Pueyo}, {Rauscher}, {Riggs}, {Ruane}, {Shaklan}, {Sirbu}, {Soummer},
  {Laurent}, \& {Zimmerman}}]{Stark2019}
Stark, C.~C., Belikov, R., Bolcar, M.~R., {et~al.} 2019, Journal of
  Astronomical Telescopes, Instruments, and Systems, 5, 024009,
  \dodoi{10.1117/1.JATIS.5.2.024009}

\bibitem[{{The HabEx Team}(2019)}]{HabEx2019}
{The HabEx Team}. 2019, {The HabEx Final Report}, Tech. rep.

\bibitem[{{The LUVOIR Team}(2019)}]{LUVOIR2019}
{The LUVOIR Team}. 2019, arXiv e-prints, arXiv:1912.06219.
\newblock \doarXiv{1912.06219}

\bibitem[{Tian {et~al.}(2014)Tian, France, Linsky, Mauas, \&
  Vieytes}]{Tian2014}
Tian, F., France, K., Linsky, J.~L., Mauas, P.~J., \& Vieytes, M.~C. 2014,
  Earth and Planetary Science Letters, \dodoi{10.1016/j.epsl.2013.10.024}

\bibitem[{Tian \& Ida(2015)}]{Tian2015}
Tian, F., \& Ida, S. 2015, Nature Geoscience, 8, 177, \dodoi{10.1038/ngeo2372}

\bibitem[{{Van Der Walt} {et~al.}(2014){Van Der Walt}, Sch{\"{o}}nberger,
  Nunez-Iglesias, Boulogne, Warner, Yager, Gouillart, \& Yu}]{VanDerWalt2014}
{Van Der Walt}, S., Sch{\"{o}}nberger, J.~L., Nunez-Iglesias, J., {et~al.}
  2014, PeerJ, 2014, e453, \dodoi{10.7717/peerj.453}

\bibitem[{{Wakeford} {et~al.}(2019){Wakeford}, {Lewis}, {Fowler}, {Bruno},
  {Wilson}, {Moran}, {Valenti}, {Batalha}, {Filippazzo}, {Bourrier},
  {H{\"o}rst}, {Lederer}, \& {de Wit}}]{Wakeford2019}
{Wakeford}, H.~R., {Lewis}, N.~K., {Fowler}, J., {et~al.} 2019, \aj, 157, 11,
  \dodoi{10.3847/1538-3881/aaf04d}

\bibitem[{Wall \& Jenkins(2003)}]{Wall2003}
Wall, J.~V., \& Jenkins, C.~R. 2003, {Practical statistics for astronomers}
  (New York :: Cambridge University Press), 277

\bibitem[{{Wang} {et~al.}(2018){Wang}, {Mawet}, {Hu}, {Ruane}, {Delorme}, \&
  {Klimovich}}]{Wang2018}
{Wang}, J., {Mawet}, D., {Hu}, R., {et~al.} 2018, Journal of Astronomical
  Telescopes, Instruments, and Systems, 4, 035001,
  \dodoi{10.1117/1.JATIS.4.3.035001}

\bibitem[{Wordsworth \& Pierrehumbert(2014)}]{Wordsworth2014}
Wordsworth, R., \& Pierrehumbert, R. 2014, Astrophysical Journal Letters, 785,
  \dodoi{10.1088/2041-8205/785/2/L20}

\bibitem[{{Wordsworth} {et~al.}(2018){Wordsworth}, {Schaefer}, \&
  {Fischer}}]{Wordsworth2018}
{Wordsworth}, R.~D., {Schaefer}, L.~K., \& {Fischer}, R.~A. 2018, \aj, 155,
  195, \dodoi{10.3847/1538-3881/aab608}

\bibitem[{Zahnle {et~al.}(2006)Zahnle, Claire, \& Catling}]{Zahnle2006}
Zahnle, K., Claire, M., \& Catling, D. 2006, Geobiology, 4, 271,
  \dodoi{10.1111/j.1472-4669.2006.00085.x}

\bibitem[{Zahnle {et~al.}(2013)Zahnle, Catling, \& Claire}]{Zahnle2013}
Zahnle, K.~J., Catling, D.~C., \& Claire, M.~W. 2013, Chemical Geology, 362,
  26, \dodoi{10.1016/j.chemgeo.2013.08.004}

\bibitem[{Zahnle {et~al.}(2019)Zahnle, Gacesa, \& Catling}]{Zahnle2019}
Zahnle, K.~J., Gacesa, M., \& Catling, D.~C. 2019, Geochimica et Cosmochimica
  Acta, 244, 56, \dodoi{10.1016/j.gca.2018.09.017}

\bibitem[{{Zhang} {et~al.}(2018){Zhang}, {Zhou}, {Rackham}, \&
  {Apai}}]{Zhang2018}
{Zhang}, Z., {Zhou}, Y., {Rackham}, B.~V., \& {Apai}, D. 2018, \aj, 156, 178,
  \dodoi{10.3847/1538-3881/aade4f}

\end{thebibliography}
\bibliographystyle{aasjournal}

\listofchanges
\end{document}